# Title: Discovering universal scaling laws in 3D printing of metals with genetic programming and dimensional analysis

**Authors:** Zhengtao Gan[1]*, Orion L. Kafka[1], Niranjan Parab[2], Cang Zhao[2], Olle Heinonen[3], Tao Sun[2], Wing Liu[1]*.

**Affiliations:**

[1] Department of Mechanical Engineering, Northwestern University.

[2] X-ray Science Division, Argonne National Laboratory.

[3] Materials Science Division, Argonne National Laboratory.

*Correspondence to: zhengtao.gan@northwestern.edu (Z. Gan), w-liu@northwestern.edu (W. K. Liu).

**Abstract:** We leverage dimensional analysis and genetic programming (a type of machine learning) to discover two strikingly simple but universal scaling laws, which remain accurate for different materials, processing conditions, and machines in metal three-dimensional (3D) printing. The first one is extracted from high-fidelity high-speed synchrotron X-ray imaging, and defines a new dimensionless number, "Keyhole number", to predict melt-pool vapor depression depth. The second predicts porosity using the Keyhole number and another dimensionless number, "normalized energy density". By reducing the dimensions of these longstanding problems, the low-dimensional scaling laws will aid process optimization and defect elimination, and potentially lead to a quantitative predictive framework for the critical issues in metal 3D printing. Moreover, the method itself is broadly applicable to a range of scientific areas.

**One Sentence Summary:** We discovered simple universal laws for keyhole depth and pore formation in metal 3D printing by using ideas from mechanics and machine learning.

**Main Text:** Metal 3D printing, sometimes called additive manufacturing, provides opportunities to create customizable parts for a variety of applications in various sectors, including transportation (*1*), space exploration (*2*), and medicine (*3*). In metal 3D printing, material is typically added layer-by-layer by local melting and (re)solidification of a precursor, usually gas-atomized metallic powders. This process provides considerable freedom to design local features, such as geometry and composition, in addition to enhancing manufacturing flexibility and reducing material waste.  However, 3D printing has a vast number of parameters to consider, with complex interactions and dependencies (*4–7*).  Many authors have quantified the impacts of various individual parameters or groups of parameters (*8–12*), but universal physical relationships, which remain valid for different materials, processing conditions, and machines, have remained elusive. The multivariate and multiphysics nature of the additive processes complicates real-time process control, parameters optimization, and materials development and selection. For example, during powder-bed fusion 3D printing, a topological depression (a so-called "keyhole") frequently forms, which is caused by vaporization recoil pressure due to intense laser-metal interactions (*13*). The geometry of the keyhole significantly affects the energy coupling mechanisms between the high-energy beam and the substrate material (*9*), which leads to unusual melt pool dynamics (*14*) and solidification defects (*3*). Although such keyholes were first studied more than 100 years ago for laser welding (*15*), high-quality in-situ experimental data were becoming available by high-speed X-ray imaging only recently (*8*). Another longstanding issue in metal 3D printing and welding is the generation of excessive porosity. Much effort has been directed at determining causal explanations for this phenomenon, and several mechanisms that lead to porosity formation have been identified, such as lack of fusion (*16*), instability of the depression zone (*13*), vaporization of volatile elements (*17*) and hydrogen precipitation (*18*). However, these efforts are still far from producing a predictive model for porosity – it is challenging to distinguish the quantitative impact of these different mechanisms on the final, observed porosity. This makes it impossible to predict the porosity type and magnitude and therefore impossible to optimize processing conditions to achieve a desired porosity in advance of processing. Low-dimensional patterns expressed as compact mathematical equations, e.g. scaling laws, provide elegant insight into the behavior of complex systems using a minimal set of parameters. This adds simplicity to otherwise highly multivariate or multi-dimensional systems and helps succinctly guide processes towards scientific discovery and engineering design (*19*). Data science and machine learning is exploring new ways of understanding high-dimensional data collected from physical processes (*20, 21*), and potentially providing new solutions to longstanding scientific challenges in 3D printing of metals. By combining dimensional analysis and genetic programming, our methodology deduces low-dimensional patterns in different aspects of 3D printing and provides simple but universal coarse-grained insight and scaling laws. The scaling laws provide causality connecting process and property, and thus enable robust real-time control and efficient process and materials optimization in 3D printing.

    We demonstrate a methodology by automatically searching for scaling laws from experimental data describing the two problems discussed above that are ubiquitous in 3D printing: keyhole depth and porosity (Fig. 1). We start by generating high-dimensional high-fidelity experimental data sets using synchrotron X-ray imaging (*8, 22, 23*) and tomography (*24*) (Fig. 1, A and B). The raw data are high-dimensional but sparse. We first perform data reduction and identify reduced forms that are invariant to the system of units using dimensional analysis (*25*) (Fig. 1C). A challenge for the dimensional analysis is requiring an unknown law to be a function of the system's states, expressed in terms of dimensionless numbers. We overcome this challenge using genetic programming (*26*) to search for explicit free-form laws underlying and embedded in the data (*20*), (Fig. 1D and Fig. S1). Low-dimensional scaling laws with the property of dimensional homogeneity can be identified and shown for keyhole depth (Fig. 1E), and for porosity (Fig. 1F).

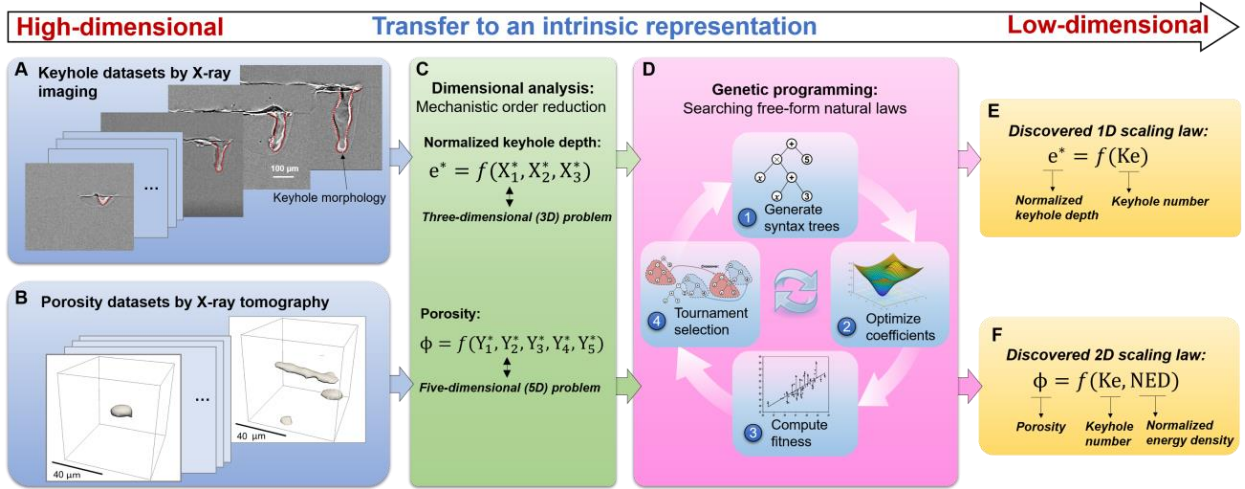

**Fig. 1. Seeking intrinsic low-dimensional and topology-preserving representations in metal 3D printing.** (**A**) X-ray imaging datasets of keyhole dynamics. (**B**) X-ray tomography datasets of porosity. (**C**) Identify reduced forms that are invariant to the system of units using dimensional analysis for the two problems. (**D**) Genetic programming is used to search for free-form natural laws that represent the dimensionless data using a minimal set of parameters. Examples of the effectiveness of the method are: (**E**) a 1D scaling law for keyhole depth, and (**F**) an intrinsic 2D dimensionless representation for porosity.

To generate keyhole data, we use high-speed synchrotron X-ray imaging data sets from three different alloys: Titanium Ti-6Al-4V (Ti64) (*8*), Aluminum 6061 (Al6061), and Stainless Steel 316 (SS316). The quantity of interest is keyhole depth $e$, which depends on seven process parameters and material properties (Fig. 2A),

$$e = f\left(\eta(P-P_0), V, d, K, \rho C_p, T_v - T_0, T_m - T_0\right) \tag{1}$$

where $\eta(P-P_0)$ is the effective laser power, $V$ is the scan speed, $d$ is the laser spot diameter, $K$ is the thermal conductivity, $\rho C_p$ is the volumetric heat capacity, $T_v - T_0$ is the relative vaporization temperature, and $T_m - T_0$ is the relative melting temperature, respectively. The absorptivity of the material is assumed to vary exponentially with keyhole depth, $\eta = \max\left(\eta_m, 0.7(1-\exp(-0.38\frac{e}{d}))\right)$, as identified with experimental measurements (*9*). The minimum absorptivity $\eta_m$ depends on the material. The minimum laser power needed to generate a keyhole is $P_0$, which can be measured from X-ray imaging (*8*) and $T_0$ is the pre-heat temperature of the substrate. The specific values of the material properties are provided in Table S1. Eight variables listed have different dimensions; however, only four are fundamental: length [L], time [T], mass [M], and temperature [K]. Thus, to make sure the physical relationship does not depend on an arbitrary choice of basic units of measurement, we use the Buckingham Pi theorem (*27*) to identify a set of four dimensionless numbers that form a definite physical relationship. Although there are multiple possible sets of dimensionless numbers, we choose $e^* = f(P^*, Pe, T^*)$, because these reflect the causality of this problem and have physical meaning: the normalized keyhole depth $e^* = \frac{e}{d}$, which is commonly used to define the keyhole aspect ratio (*11*); the normalized power $P^* = \frac{\eta(P-P_0)}{dK(T_v - T_0)}$, which is the ratio of the effective laser power to another power only related to material properties; the Peclet number $\text{Pe} = \dfrac{V \rho C_p d}{K}$, which is a well-known dimensionless number representing the ratio of heat storage rate to heat conductive transport rate (*12*); and the normalized temperature $\text{T}^* = \dfrac{T_v - T_0}{T_m - T_0}$, which is the ratio of the relative vaporization point to the relative melting point.

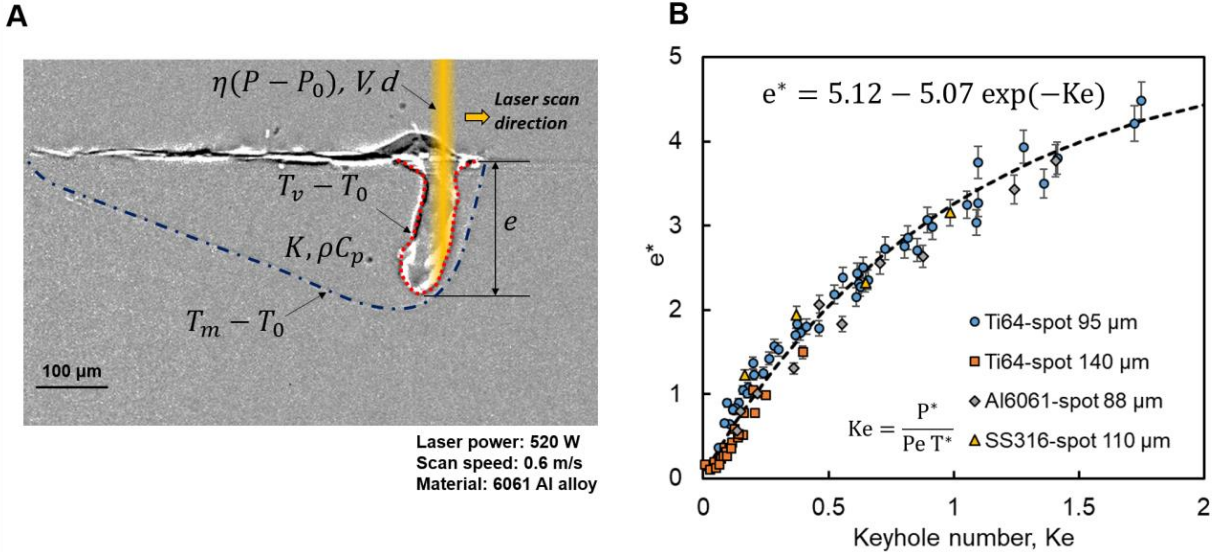

**Fig. 2. Discovering a one-dimensional pattern controlled by the Keyhole number for keyhole depth.** (**A**) Illustrative X-ray imaging result, showing a snapshot in time of the cross-section of a melt pool with phase density contrast. A schematic of the estimated laser beam profile is artificially superimposed (yellow). The keyhole boundary is outlined with a red dotted line, and the melt pool boundary is outlined with a blue dash-dotted line. The keyhole depth is initially assumed to depend on seven process parameters and material properties. (**B**) Identified scaling law for keyhole depth, which depends on only one parameter Ke, and is *universal* -- the relationship is valid for different materials and processing conditions as indicated by the different colored points.

We find the exact form and coefficients of the unknown function f(•) by using genetic programming (*26*) based on evolutionary computation to search the space of mathematical expressions for a relationship that minimize various error metrics (Fig. S1). We provide a detailed description of data separation and the procedural setup of the genetic programming algorithm in Fig. S2. Our method identifies five thousand candidate mathematical expressions in the final evolutionary generation of the genetic programming (Fig. S2D). These mathematical expressions, or models, have different accuracy and expressional complexity (detailed definitions of these terms are presented in Supplementary Materials). A parsimonious model is

$$e^* = 5.12 - 5.07 \exp\left(-\dfrac{1.0\,P^*}{\text{PeT}^*}\right) \qquad (2)$$

which minimizes the number of parameters or variables but maximizes explanatory predictive power. A comparison between the best model and other candidate models is presented in Fig. S3. Surprisingly, although the parsimonious model involves all three independent input variables, by using the evolutionary search we find that grouping them together best matches the data. This implies that these inputs might not be independent, but that only certain combinations of them matter. Thus, we define a new dimensionless number, the Keyhole number Ke, as the combination of $P^*$, Pe and $T^*$: $Ke = \dfrac{P^*}{PeT^*}$, such that the normalize keyhole depth only depends on the Keyhole number (Fig. 2B). All the data lie on a single curve, even though they correspond to various laser power, scan speed, laser spot size, and material. Thus, our Keyhole number Ke exhibits *universal scaling* for keyhole depth (Table S2 and Fig. S4). The Keyhole number,

$$Ke = \left(\frac{\eta(P-P_0)}{Vd^2}\right)\left(\frac{1}{\rho C_p(T_v-T_0)}\right)\left(\frac{T_m-T_0}{T_v-T_0}\right) \qquad (3)$$

represents a product of the effective energy density of the laser heat source, the inverse of the sensible heat of vaporization, and a dimensionless coefficient representing the relative distance between the vaporization and melting isotherms. The machine learning-based search confirms the Keyhole number is the minimal representation of the keyhole depth. Note that the exponential behavior of the scaling law (Eq. 2) stems from the exponential law of absorptivity with the keyhole depth (*9*). We expect that the scaling law for keyhole depth will transition to a linear relation when the Keyhole number is beyond a threshold where the absorptivity approaches one.

In order to demonstrate the generality of our methodology, we explore another even more challenging problem: porosity in 3D printing. Using the same paradigm described above, the porosity problem starts with a nine-dimensional (9D) description,

$$\phi = f\left(\eta P, V, d, K, \rho C_p, T_v - T_0, T_m - T_0, H, L\right) \qquad (4)$$

where $\phi$ describes the porosity. In this case, the process is a multi-track and multi-layer one, so we add two more length scales: hatch spacing $H$ and layer thickness $L$. It is intractable to directly identify a suitable 9D relationship because millions to billions of data points would be required. Using the first step of our process, dimensional analysis, we deduce an unknown function of five dimensionless variables, $\phi = f(P_m^*, Pe, P_v^*, H^*, T^*)$. These variables are melting efficiency $P_m^* = \dfrac{\eta P}{LK(T_m-T_0)}$, Peclet number $Pe = \dfrac{V\rho C_p d}{K}$, vaporization efficiency $P_v^* = \dfrac{\eta P}{HK(T_v-T_0)}$, normalized hatch spacing $H^* = \dfrac{H}{d}$, and normalized temperature $T^* = \dfrac{T_v-T_0}{T_m-T_0}$.

In the next step, using genetic programming to identify a parsimonious model, we find ten thousand possible mathematical expressions with different accuracy and expressional complexity in the final evolutionary generation of the genetic program (Fig. 3). The Pareto frontier is marked with a dashed black curve, and we have one extremum (highlighted with a red circle) on the Pareto frontier. This model is

$$\phi = 0.0591\,\text{erfc}\left(\frac{P_m^*}{PeH^*} - 2.04\right) + 0.0591\,\text{erf}\left(\frac{P_v^*}{PeT^*} - 2.09\right) + 0.0639 \qquad (5)$$

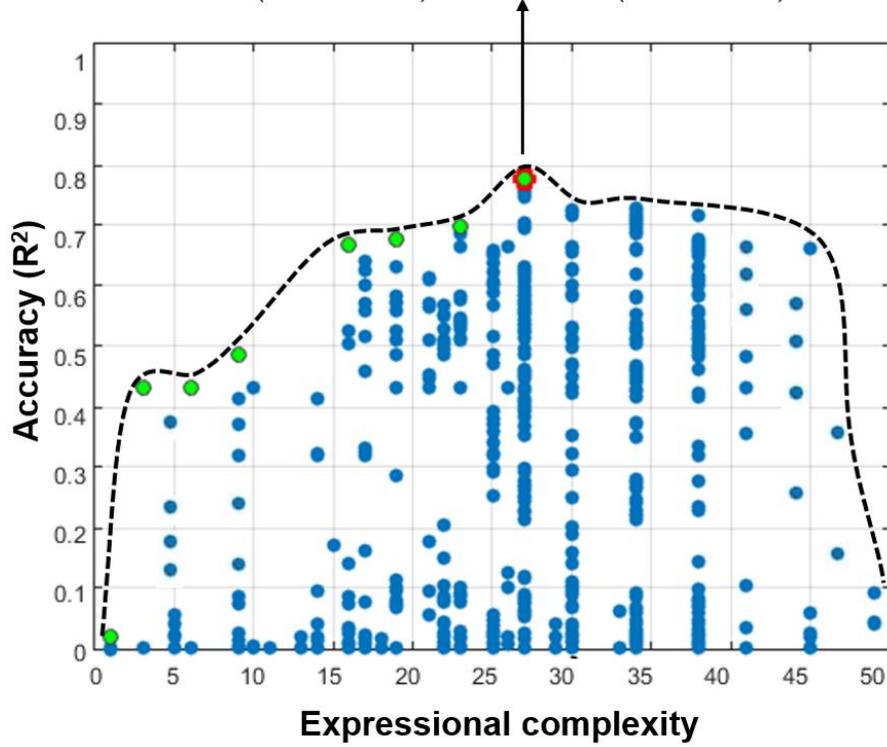

**Fig. 3. Searching for a parsimonious model for porosity.** The figure plots accuracy vs. complexity of 10,000 candidate models at the final generation of genetic programming. The Pareto frontier for candidate models is marked with a dashed black curve. One extremum can be found on the Pareto frontier, highlighted in red. The detailed expression of this parsimonious model is given above.

Details regarding setup are given in Fig. S5. There are two interesting facets to the expression for the parsimonious model. First, the machine learning-based search identifies only one function, the Gaussian error function $\text{erf}(x) = \frac{1}{\sqrt{\pi}} \int_{-x}^{x} e^{-t^2} dt$, that completely describes the data, although other eight functions are optional in the function set (Fig. S5B), and $\text{erfc}(x)$ is the complementary error function defined as $\text{erfc}(x) = 1 - \text{erf}(x)$. This implies that the origin of porosity formation is related to specific Gaussian random processes. Second, the five input variables are grouped into two dimensionless numbers. We interpret the first dimensionless number as a "normalized energy density" ( NED ), defined by

$$\text{NED} = \frac{P_m^*}{\text{PeH}^*} = \left(\frac{\eta P}{VHL}\right)\left(\frac{1}{\rho C_p (T_m - T_0)}\right) \quad (6)$$

and representing the ratio of effective energy density to sensible heat of melting. Interestingly, the Keyhole number

$$\text{Ke} = \frac{P_v^*}{\text{PeT}^*} = \left(\frac{\eta P}{VdH}\right)\left(\frac{1}{\rho C_p (T_v - T_0)}\right)\left(\frac{T_m - T_0}{T_v - T_0}\right) \tag{7}$$

appears again as the second independent parameter, although this model for the porosity is entirely independent from the previous search that also identified the Keyhole number as a governing dimensionless parameter. In this case, the Keyhole number includes a product of hatch spacing $H$ and laser spot diameter $d$ to account for multiple tracks, whereas for a single track process (as used for the first Keyhole number demonstration) the square of laser spot diameter $d^2$ is sufficient. The Keyhole number includes $\eta P$ rather than $\eta(P - P_0)$ as used for the first case since the values of $P_0$ (the minimum laser power needed to generate a keyhole) are not available in the porosity case. Also note that the Keyhole number does not include the layer thickness of powder bed $L$, implying the keyhole depths is independent of the existence of the powder bed; this is consistent with in-situ experimental observations (*8*). The appearance of the Keyhole number in these seemingly unrelated problems (keyhole geometry and porosity) implies the mechanism for porosity formation would be similar to keyhole formation and would not appear below a critical Keyhole number. The proposed method provides a new avenue to quantify this similarity.

Our newly discovered two-dimensional pattern (scaling law) for porosity in 3D printing includes two terms and a constant (Fig. 4). The first term of the scaling law is related only to the NED, depicted in Fig. 4A. This first term fits data with low NED, i.e. less than around five in this case, which is a regime characterized by lack of fusion defects. When NED exceeds that critical value, lack of fusion porosity is avoided and the efficacy of NED as a predictor of porosity decreases. Simultaneously, the second term and the constant part of the scaling law begin to represent the observed data well, as shown in Fig. 4B. Physically, this is explained by higher input energies resulting in vaporization porosity, which can be characterized by the Keyhole number. The combination of the two mechanisms as a 2D scaling law is visualized (shown as a mesh surface) along with experimental data points (colored dots) in Fig. 4C. The model only depends on two dimensionless numbers, and scales well with various process parameters, materials, and machines; the data shown are in fact from three independent sources (*12, 16, 28*). Our method is effective at discovering dominant low-dimensional representations of physical systems, given enough data. This could benefit disciplines ranging from biology to material science, where governing laws are elusive despite abundant data.

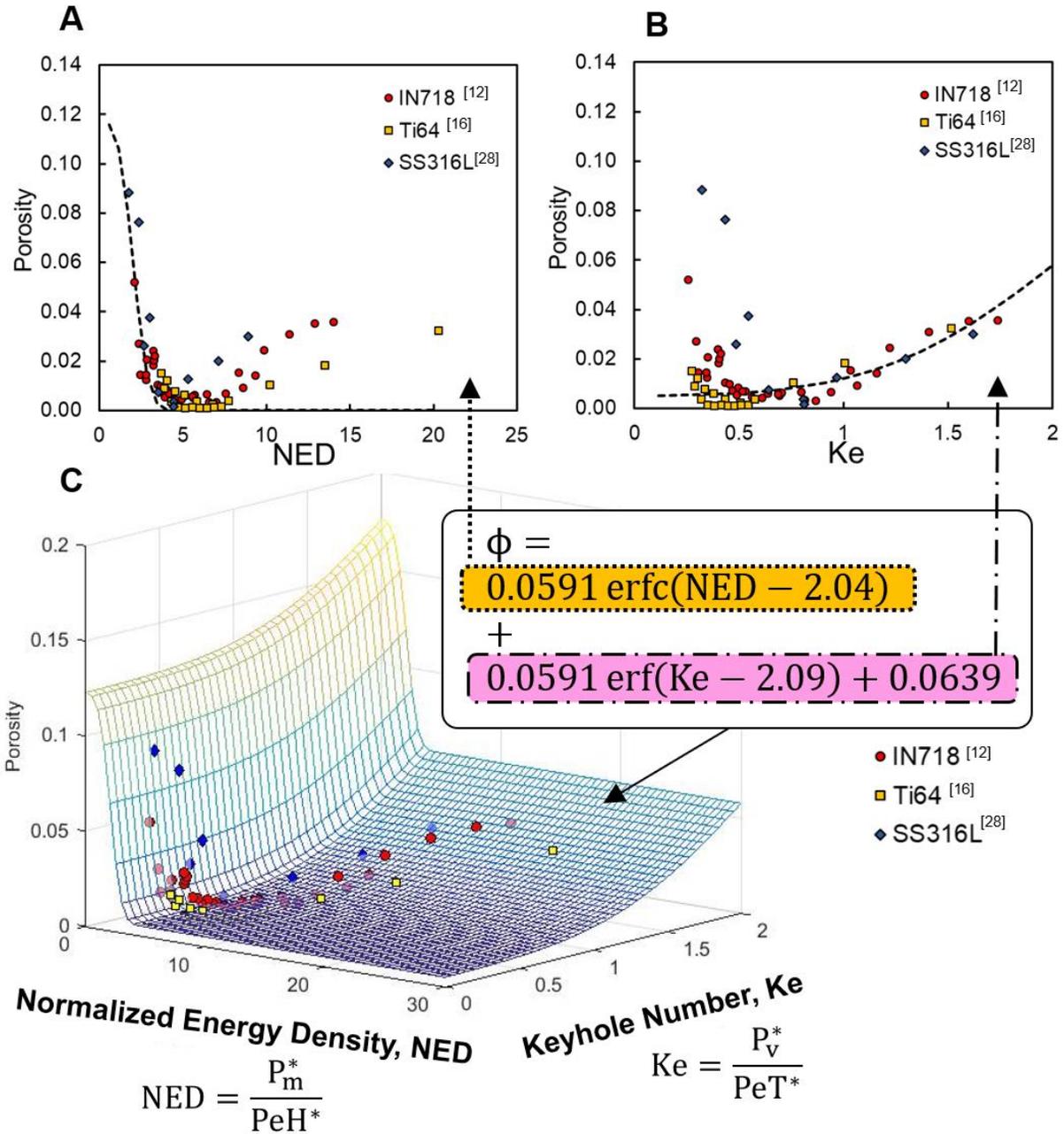

**Fig. 4. Two-dimensional scaling law for porosity.** (**A**) The first term of the scaling law with respect to normalized energy density NED, which quantifies porosity formed via the lack of fusion process. (**B**) The second and the constant term of the scaling law with respect to the Keyhole number, Ke, which quantifies the vaporization porosity. (**C**) Two-dimensional scaling law with respect to normalized energy density NED and the Keyhole number Ke, with a surface formed by the combination of the two parameters. This reduced function space is much more easily visualized and interpreted than the original 9D space.

**References and Notes:**

1. P. Vora, K. Mumtaz, I. Todd, N. Hopkinson, AlSi12 in-situ alloy formation and residual stress reduction using anchorless selective laser melting. *Addit. Manuf.* **7**,12-19 (2015).

2. Z. C. Eckel, C. Zhou, J. H. Martin, A. J. Jacobsen, W. B. Carter, T. A. Schaedler, Additive manufacturing of polymer-derived ceramics. *Sci.* **351**, 58-62 (2016).

**Acknowledgments:** We thank the Center for Hierarchical Materials Design (CHiMaD), in particular, our ongoing work with Lyle E. Levine has been synergistic. **Funding:** W.K.L., Z.G., and O.L.K. were supported by the National Science Foundation (NSF) through grant CMMI-1762035. O.L.K. acknowledges support through the NSF Graduate Research Fellowship under Grant No. DGE-1324585. N.D.P., C.Z., and T.S. would like to acknowledge Laboratory Directed Research and Development (LDRD) funding from Argonne National Laboratory, provided by the Director, Office of Science, of the U.S. Department of Energy under Contract No. DE-AC02-06CH11357; the work by O.G.H. was performed under financial assistance award 70NANB14H012 from U.S. Department of Commerce, National Institute of Standards and Technology as part of the CHiMaD. This research used resources of the Advanced Photon Source, a U.S. Department of Energy (DOE) Office of Science User Facility operated for the DOE Office of Science by Argonne National Laboratory under Contract No. DE-AC02-06CH11357. Some of the results shown in this paper were used in writing the to-be-funded NSF proposal CMMI-1934367 (starting date Dec 1, 2019). **Author contributions:** O.G.H. and W.K.L. procured funding and supervised the project. Z.G., O.L.K. and O.G.H. analyzed the results and wrote the manuscript. N.D.P., C.Z. and T.S. designed and conducted the experiments. Z.G. implemented the genetic programming and dimensional analysis; **Competing interests:** Authors declare no competing interests. **Data and materials availability:** All data is available in the main text or the supplementary materials.

**Supplementary Materials:**

Materials and Methods

Figures S1-S6

Tables S1-S2

Movies S1-S11

References (29,*30*)

# Supplementary Materials for

## Discovering universal scaling laws in 3D printing of metals with genetic programming and dimensional analysis

Zhengtao Gan, Orion L. Kafka, Niranjan Parab, Cang Zhao, Olle Heinonen, Tao Sun, Wing Liu.

Correspondence to: zhengtao.gan@northwestern.edu, w-liu@northwestern.edu.

**This PDF file includes:**

Materials and Methods

Figs. S1 to S6

Tables S1 to S2

Captions for Movies S1 to S11

**Other Supplementary Materials for this manuscript include the following:**

Movies S1 to S11

**Materials and Methods**

Materials

We manufactured 50 mm by 3 mm by 0.75 mm samples of aluminum (Al6061, McMaster-Carr, USA) and 50 mm by 3 mm by 0.5 mm samples of stainless steel (SS316, McMaster-Carr, USA) from as-received plates using conventional manufacturing methods (cutting and milling).

Selective laser melting apparatus

We built a custom selective laser melting apparatus by combining a ytterbium fiber laser (YLR-500-AC, IPG, USA), a galvo laser scanner (IntelliSCANde30, SCANLAB GmbH., Germany) and a vacuum chamber. The laser wavelength was 1070 nm and the maximum power was 540 W. We operated it in single mode, providing a Gaussian beam profile. At the focal plane, the laser spot size was ≈ 56 μm. In this study, samples were positioned at a certain distance below the focal plane to achieve larger laser spot sizes (1.5 mm for a spot size of 88 μm, 1.8 mm for 95 μm, 2.5 mm for 110 μm, and 3.5 mm for 140 μm). Single track laser melting experiments were performed on the samples under various laser powers (208 W to 520 W) and scan speeds (0.3 m/s to 1.2 m/s). More details of this apparatus are provided elsewhere (*22, 23*).

High-speed X-ray imaging

The high-speed X-ray imaging experiments were performed at beam line 32-ID-B of the Advanced Photon Source (APS) at Argonne National Laboratory. A short period (18 mm) undulator with the gap set to 12 mm was used to generate polychromatic X-rays with the first harmonic energy centered at 24.7 keV. The X-ray photons were allowed to pass through the sample while the laser was traversing on the top surface (50 mm by 1 mm area for aluminum and 50 mm by 500 μm area for stainless steel samples). The propagated X-ray photons were converted to visible light using a LuAG:Ce scintillator (10 mm diameter, 100 μm thickness) and recorded with a high-speed camera (Photron FastCam SA-Z, USA) after passing a 45° reflection mirror, a relay lens, and a 10× objective lens. The nominal spatial resolution of the imaging system was 1.93 μm/pixel. We recorded high-speed X-ray image at frame rates between 20,000 and 50,000 frames per second with exposure times between 1 μs to 40 μs, with higher exposure times used for stainless steel samples. The detector-to-sample distance was between 300 mm and 400 mm. A series of delay generators were used to trigger the X-ray shutters, laser systems, and camera sequentially. More details of the high-speed X-ray imaging setup are provided in Parab et al. (*23*) and Zhao et al. (*22*).

Image processing and quantification

The images of keyholes and melt pools were processed and analyzed using ImageJ (*29*). Each image stack includes a time series of images. For each we first we duplicated the original image stack (A) to create an identical stack (B). Second, we duplicated the first slice of stack A and the last slice of the stack B. Third, we divided image stack A pixel-wise for each slice by stack B to reduce background noise and increase contrast of the images. Fourth, we used a despeckling median filter to further eliminate speckle noise. Finally, we adjusted the brightness and contrast to enhance contrast and ease visualization and size measurement. We quantified the size of the melt pool and keyhole based on contours of attenuation contrast. This involved manual inspection and measurement of each image in all the image stacks in ImageJ, recording the XY coordinates describing length and depth of the melt pool and keyhole for each image. The maximum, minimum, mean and standard deviation for each stack were calculated.

Genetic programming

     We implemented genetic programming using an open-source MATLAB-based software platform: GPTIPS 2 (*30*). As shown in Fig. S1, each run followed a series of steps. First, we determined a terminal set and a function set (Fig. S2B and S5B). Then we randomly created an initial population of candidate models using a tree structure (Fig. S1B). Second, the coefficients of each candidate model were optimized using the stochastic gradient decent (SGD) method. Third, a fitness score, i.e. the root-mean-square error (RMSE) and the expressional complexity of each candidate model was calculated. The expressional complexity was computed by summing the node count of each syntax tree and all its possible full sub-trees. Forth, based on the fitness and complexity, Pareto tournament selection, crossover, and mutation operators (Fig. S1, B and C) were conducted to create the population in the next generation. We repeated the above steps until a predetermined number of final generations was reached (in this case 30). To determine a parsimonious model, we used the Akaike information criterion (AIC) and the Bayesian information criterion (BIC) based on the computed coefficient of determination, $R^2$, and the expressional complexity of each model at the final generation. Multiple (five for the keyhole case and ten for the more complex porosity case) independent runs were conducted to ensure that the algorithm converged to the same parsimonious model. To facilitate estimation of the predictive power of our models, the initial datasets were divided into training data and test data. The training data was used to train the genetic programming algorithm, and the test data was used to test the generalization capability of discovered model. For the keyhole problem, we used Ti6Al4V (*8*) data for training, and the combined Al6061 and SS316 data for testing (Fig. S2A). For the porosity problem, we used the Inconel 718 (*12*) and Ti6Al4V (*16*) data for training, and the SS316L (*28*) data for testing (Fig. S5A).

Dimensional analysis

     Derivations of the dimensional analysis for the keyhole depth and porosity are shown in Fig. S6. The thermophysical properties at the melting point were used in the dimensional analysis (Table S1). Note that other dimensionless variables that are the combination of the proposed ones might also be obtained. Any choice that includes a complete set of the dimensionless variables (all dimensional governing variables are included) is equivalent and will not affect the final results of the genetic programming. However, we chose sets that include easily physically interpreted dimensionless variables.

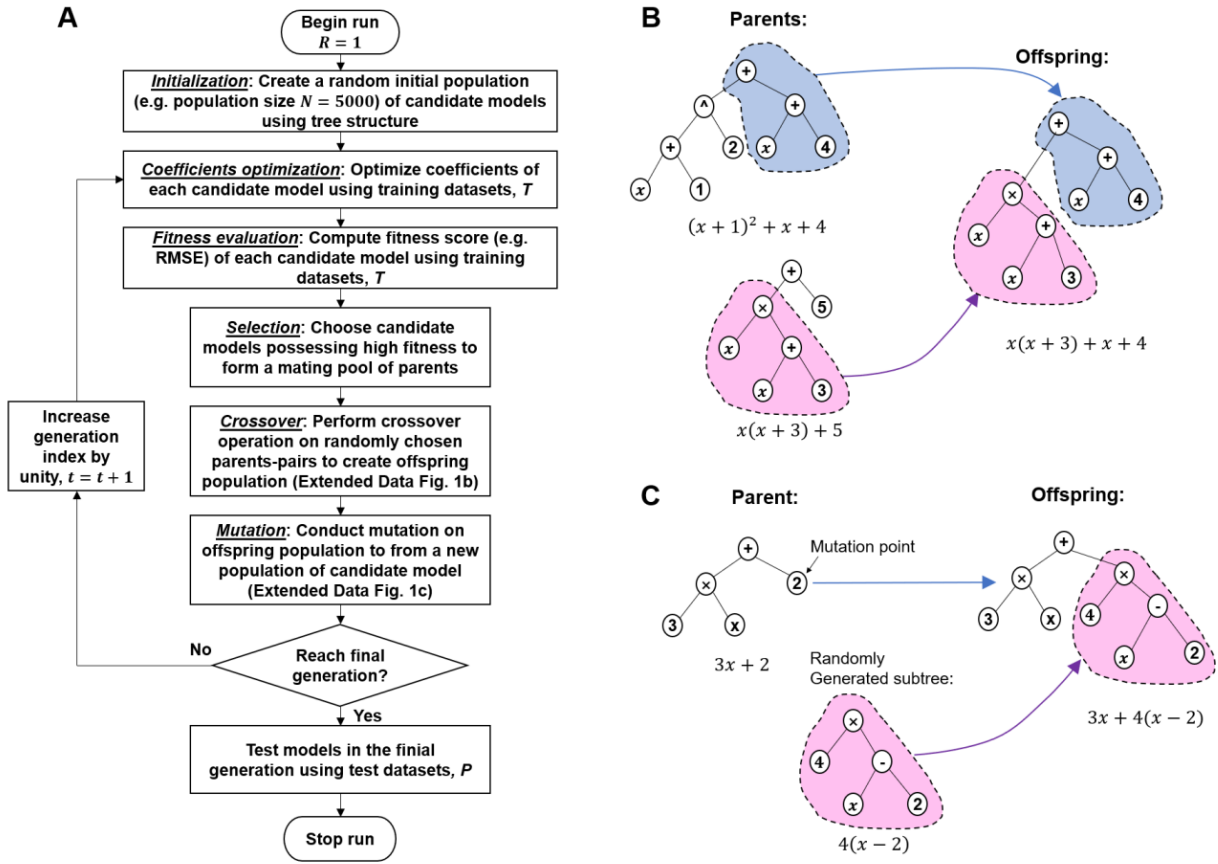

**Fig. S1. Genetic programming implementation.** (**A**) Flowchart of genetic programming. (**B**) Example of crossover operator. (**C**) Example of mutation operator.

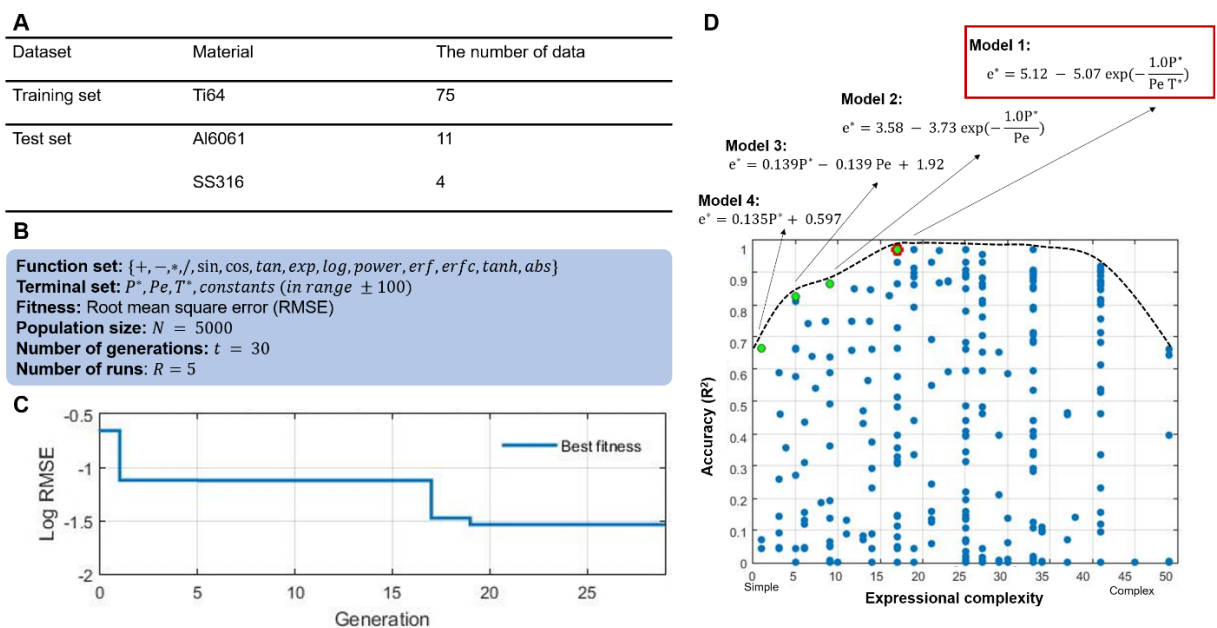

**Fig. S2. Data separation and procedural setup of the genetic programming algorithm for keyhole depth problem.** (**A**) Data separation. (**B**) Procedural setup Implemented. (**C**) Convergence curve of the first run with various generation. (**D**) Candidate models at the final generation depicted in terms of accuracy vs. complexity. The Pareto front for candidate models is marked with a dashed black curve. Model 1 at the Pareto cliff corresponds to the discovered model with the best trade-off between accuracy and complexity (highlighted with a red box in the figure).

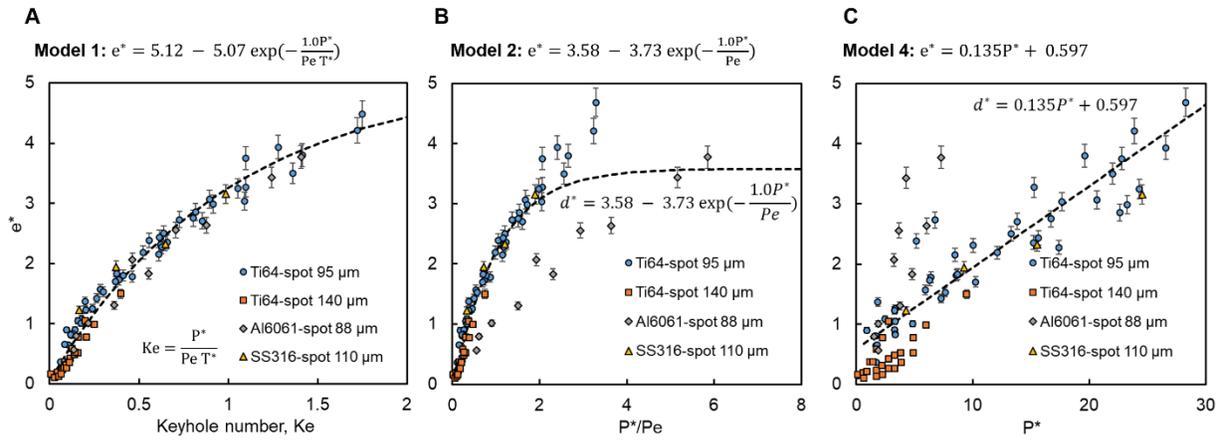

**Fig. S3. Candidate models at the final generation.** (**A**) Model 1 including three input variables. (**B**) Model 2 including two input variables. (**C**) Model 4 including only one input variable.

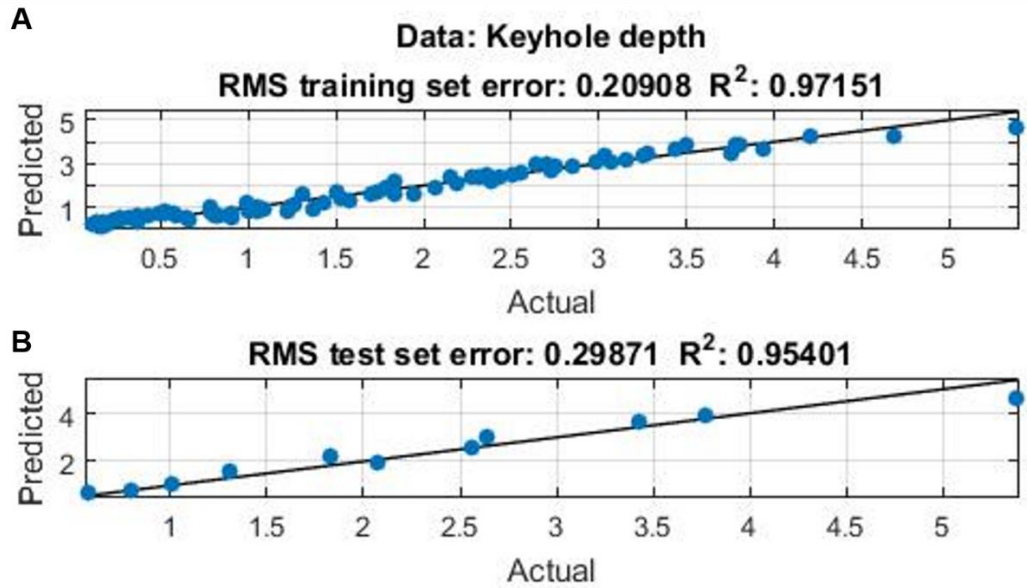

**Fig. S4. Performance of the minimal model.** (**A**) Root mean square (RMS) error in training dataset. (**B**) RMS error in test dataset.

## A

| Dataset | Material | The number of data |
|---|---|---|
| Training set | Inconel 718 | 35 |
| | Ti6Al4V | 19 |
| Test set | SS316L | 11 |

## B

**Function set:** $\{+, -, *, /, \sin, \cos, \tan, \exp, \log, power, erf, erfc, \tanh, abs\}$
**Terminal set:** $P_m^*, Pe, P_v^*, \Delta_h^*, T^*, constants\ (in\ range\ \pm 100)$
**Fitness:** RMSE
**Population size:** $N = 10,000$
**Number of generations:** $t = 30$
**Number of runs:** $R = 10$

## C

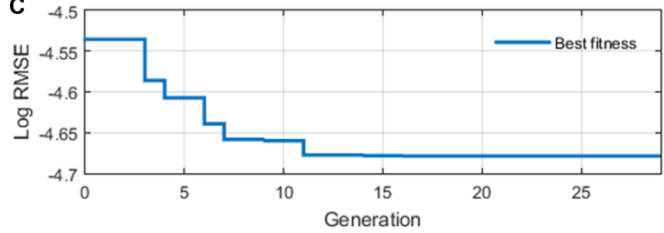

## D

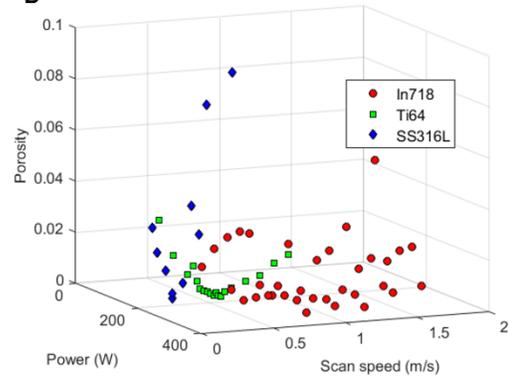

**Fig. S5. Data separation and procedural setup of the genetic programming algorithm for porosity problem.** (**A**) Data separation. (**B**) Procedural setup. (**C**) Convergence curve of the first run with various generation. (**D**) Visualization of datasets using laser power and scan speed as coordinates.

**A**

The keyhole depth $e$ depends on the following governing variables:
$$e = f(\eta(P - P_0), V, d, K, \rho C_p, T_v - T_0, T_m - T_0)$$
The corresponding dimensions of them:

$e$   $\eta(P-P_0)$   $V$   $d$   $K$   $\rho C_p$   $T_v - T_0$   $T_m - T_0$

$[L]$   $[\frac{ML^2}{T^3}]$   $[\frac{L}{T}]$   $[L]$   $[\frac{ML}{T^3K}]$   $[\frac{M}{LT^2K}]$   $[K]$   $[K]$

Four independent dimensions: $[L], [T], [M], [K]$

Four dimensionless variables:

$$e^* = \frac{e}{d} \qquad Pe = \frac{V\rho C_p d}{K}$$

$$P^* = \frac{(P-P_0)\eta}{dK(T_v - T_0)} \qquad T^* = \frac{T_v - T_0}{T_m - T_0}$$

**B**

The porosity $\phi$ depends on the following governing variables:
$$\phi = f(\eta P, V, d, K, \rho C_p, T_v - T_0, T_m - T_0, H, L)$$
The corresponding dimensions of them:

$\phi$   $\eta P$   $V$   $d$   $K$   $\rho C_p$   $T_v - T_0$   $T_m - T_0$   $H$   $L$

$[1]$   $[\frac{ML^2}{T^3}]$   $[\frac{L}{T}]$   $[L]$   $[\frac{ML}{T^3K}]$   $[\frac{M}{LT^2K}]$   $[K]$   $[K]$   $[L]$   $[L]$

Four independent dimensions: $[L], [T], [M], [K]$

Six dimensionless variables:

$$\phi \qquad Pe = \frac{V\rho C_p d}{K} \qquad H^* = \frac{H}{d}$$

$$P_m^* = \frac{P\eta}{LK(T_m - T_0)} \qquad P_v^* = \frac{P\eta}{HK(T_v - T_0)} \qquad T^* = \frac{T_v - T_0}{T_m - T_0}$$

**Fig. S6. Derivation of dimensional analysis.** (**A**) For keyhole depth problem. (**B**) For porosity problem.

| Properties & units | Ti64 | Al6061 | SS316 |
|---|---|---|---|
| Density (liquid) $\rho$ ($\frac{kg}{m^3}$) | 5316 | 2415 | 6681 |
| Heat capacity (liquid) $C_p$ ($\frac{J}{kg \cdot K}$) | 730 | 1170 | 790 |
| Thermal conductivity (liquid) $K$ ($\frac{W}{m \cdot K}$) | 30 | 90 | 26.9 |
| Melting point $T_m$ ($K$) | 1900 | 894 | 1723 |
| Vaporization point $T_v$ ($K$) | 3315 | 2792 | 3122 |
| Minimal absorptivity $\eta_m$ | 0.26 | 0.1 | 0.33 |
| Minimum laser power needed to generate a keyhole $P_0$ ($W$) | 100 for spot 95μm<br>125 for spot 140μm | 250 for spot 88μm | 100 for spot 110μm |

**Table S1. Thermophysical properties of the employed materials.**

| X-ray imaging result | Laser power, $P$ (W) | Scan speed, $V$ (m/s) | Keyhole number, $Ke$ | Mean normalized keyhole depth, $e^*$ |
|---|---|---|---|---|
| 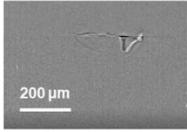 | 416 | 0.9 | 0.15 | 0.79 |
| 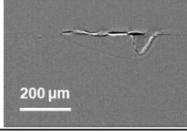 | 520 | 0.9 | 0.36 | 1.31 |
| 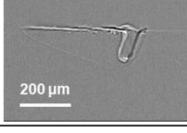 | 520 | 0.75 | 0.56 | 1.83 |
| 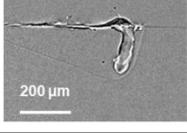 | 520 | 0.6 | 0.88 | 2.64 |
| 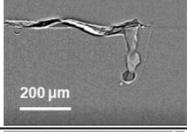 | 416 | 0.3 | 1.24 | 3.43 |
| 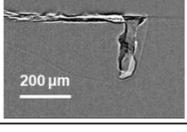 | 520 | 0.45 | 1.41 | 3.77 |

**Table S2. Illustrative X-ray imaging result, process parameters, Keyhole number and normalized keyhole depth of several demonstrative cases of Al6061.**

**Movie S1 to S11.**

**High-speed X-ray imaging of stationary and scanning laser melting of Al6061 – representative movies**

Movie S1: Progression of the melt pool and vapor depression in Al6061 bare plate under stationary laser illumination. The sample thickness is 0.75 mm. The imaging frame rate is 50,000 fps. The laser spot size is 88 µm, the power is 416 W, the scan speed is 0.3 m/s. A pixel resolution of 1.98 um. The exposure time for each image is 1 us.

Movie S2: Progression of the melt pool and vapor depression in Al6061 bare plate under stationary laser illumination. The sample thickness is 0.75 mm. The imaging frame rate is 50,000 fps. The laser spot size is 88 µm, the power is 416 W, the scan speed is 0.6 m/s. A pixel resolution of 1.98 um. The exposure time for each image is 1 us.

Movie S3: Progression of the melt pool and vapor depression in Al6061 bare plate under stationary laser illumination. The sample thickness is 0.75 mm. The imaging frame rate is 50,000 fps. The laser spot size is 88 µm, the power is 416 W, the scan speed is 0.9 m/s. A pixel resolution of 1.98 um. The exposure time for each image is 1 us.

Movie S4: Progression of the melt pool and vapor depression in Al6061 bare plate under stationary laser illumination. The sample thickness is 0.75 mm. The imaging frame rate is 50,000 fps. The laser spot size is 88 µm, the power is 416 W, the scan speed is 0.75 m/s. A pixel resolution of 1.98 um. The exposure time for each image is 1 us.

Movie S5: Progression of the melt pool and vapor depression in Al6061 bare plate under stationary laser illumination. The sample thickness is 0.75 mm. The imaging frame rate is 50,000 fps. The laser spot size is 88 µm, the power is 416 W, the scan speed is 0.45 m/s. A pixel resolution of 1.98 um. The exposure time for each image is 1 us.

Movie S6: Progression of the melt pool and vapor depression in Al6061 bare plate under stationary laser illumination. The sample thickness is 0.75 mm. The imaging frame rate is 50,000 fps. The laser spot size is 88 µm, the power is 520 W, the scan speed is 0.3 m/s. A pixel resolution of 1.98 um. The exposure time for each image is 1 us.

Movie S7: Progression of the melt pool and vapor depression in Al6061 bare plate under stationary laser illumination. The sample thickness is 0.75 mm. The imaging frame rate is 50,000 fps. The laser spot size is 88 µm, the power is 520 W, the scan speed is 0.45 m/s. A pixel resolution of 1.98 um. The exposure time for each image is 1 us.

Movie S8: Progression of the melt pool and vapor depression in Al6061 bare plate under stationary laser illumination. The sample thickness is 0.75 mm. The imaging frame rate is 50,000 fps. The laser spot size is 88 µm, the power is 520 W, the scan speed is 0.6 m/s. A pixel resolution of 1.98 um. The exposure time for each image is 1 us.

Movie S9: Progression of the melt pool and vapor depression in Al6061 bare plate under stationary laser illumination. The sample thickness is 0.75 mm. The imaging frame rate is 50,000 fps. The laser spot size is 88 µm, the power is 520 W, the scan speed is 0.9 m/s. A pixel resolution of 1.98 um. The exposure time for each image is 1 us.

Movie S10: Progression of the melt pool and vapor depression in Al6061 bare plate under stationary laser illumination. The sample thickness is 0.75 mm. The imaging frame rate is 50,000 fps. The laser spot size is 88 µm, the power is 520 W, the scan speed is 1.2 m/s. A pixel resolution of 1.98 um. The exposure time for each image is 1 us.

Movie S11: Progression of the melt pool and vapor depression in Al6061 bare plate under stationary laser illumination. The sample thickness is 0.75 mm. The imaging frame rate is 50,000 fps. The laser spot size is 88 µm, the power is 520 W, the scan speed is 0.75 m/s. A pixel resolution of 1.98 um. The exposure time for each image is 1 us.